\documentstyle[12pt,amssymb]{article}

\def\fnote#1#2{\begingroup\def\thefootnote{#1}\footnote{#2}\addtocounter{footnote}{-1}\endgroup}

\def\inbar{\vrule height1.5ex width.4pt depth0pt}
\def\IB{\relax{\rm I\kern-.18em B}}
\def\IC{\relax\,\hbox{$\inbar\kern-.3em{\rm C}$}}
\def\ID{\relax{\rm I\kern-.18em D}}
\def\IE{\relax{\rm I\kern-.18em E}}
\def\IF{\relax{\rm I\kern-.18em F}}
\def\IG{\relax\,\hbox{$\inbar\kern-.3em{\rm G}$}}
\def\IH{\relax{\rm I\kern-.18em H}}
\def\II{\relax{\rm I\kern-.18em I}}
\def\IK{\relax{\rm I\kern-.18em K}}
\def\IL{\relax{\rm I\kern-.18em L}}
\def\IM{\relax{\rm I\kern-.18em M}}
\def\IN{\relax{\rm I\kern-.18em N}}
\def\IO{\relax\,\hbox{$\inbar\kern-.3em{\rm O}$}}
\def\IP{\relax{\rm I\kern-.18em P}}
\def\IQ{\relax\,\hbox{$\inbar\kern-.3em{\rm Q}$}}
\def\IR{\relax{\rm I\kern-.18em R}}
\def\IT{\relax{\rm I\kern-.18em T}}
\def\ZZ{\relax{\sf Z\kern-.4em Z}}

\def\a{\alpha}   \def\b{\beta}      
      
  \def\om{\omega}   \def\si{\sigma}

\def\cA{{\cal A}} \def\cB{{\cal B}}
   
 \def\cH{{\cal H}} \def\cI{{\cal I}} 
   
\def\cO{{\cal O}} \def\cP{{\cal P}}

\def\pfrak{{\mathfrak p}}

\def\mathC{{\mathbb C}}  \def\mathN{{\mathbb N}}
\def\mathP{{\mathbb P}}  \def\mathQ{{\mathbb Q}}  \def\mathZ{{\mathbb Z}}

  \def\bx{{\bar x}} \def\by{{\bar y}}

\def\Ghat{{\hat G}}

\def\hurmG{{\hat{\urmG}}}

\def\fnote#1#2{\begingroup\def\thefootnote{#1}\footnote{#2}\addtocounter
{footnote}{-1}\endgroup}
\def\beq{\begin{equation}}
\def\eeq{\end{equation}}
\def\bea{\begin{eqnarray}}
\def\eea{\end{eqnarray}}

\def\lleq#1{\label{#1}\eeq}
\let\nn=\nonumber
\def\tabroom{\hbox to0pt{\phantom{\Huge A}\hss}}
\def\notin{\ \hbox{{$\in$}\kern-.51em\hbox{/}}}

\def\lra{\longrightarrow}
\def\lolra{{\longleftrightarrow}}
\def\ra{{\rightarrow}}

\def\vphi{\varphi}

  \def\E1Fq{E_1/\IF_q}

\def\uhG{{\underline \Ghat}}  \def\urmG{{\underline \rmG}}

\def\urmG{{\underline \rmG}}

\def\IFq{{\IF_q}}  

\def\XIFp{{X/\IF_p}}  \def\XIFpr{{X/\IF_{p^r}}}

\def\rmG{{\rm G}} \def\rmH{{\rm H}} \def\rmL{{\rm L}}

\def\rmgcd{{\rm gcd}} \def\rmord{{\rm ord}}
\def\rmmod{{\rm mod}} \def\rmsign{{\rm sign}}

\def\rmGal{{\rm Gal}}  \def\rmGSO{{\rm GSO}}

\def\rmHW{{\rm HW}}   \def\rmRe{{\rm Re}} \def\rmSL{{\rm SL}}
\def\rmSU{{\rm SU}}

\def\notdiv{{\relax{~|\kern-.34em /~}}}

\begin{document}

\parindent=0pt


\hfill {\bf NSF-KITP--04--105}

\vskip 0.9truein

\centerline{\Large {\bf Geometric Kac-Moody Modularity}}

\vskip .5truein

\centerline{\sc Monika Lynker$^1$\fnote{\diamond}{email:
mlynker@iusb.edu} and Rolf Schimmrigk$^2$\fnote{\dagger}{email:
netahu@yahoo.com, rschimmr@kennesaw.edu}}

\vskip .2truein

\centerline{\it $^1$Indiana University South Bend} \vskip
.05truein \centerline{\it 1700 Mishawaka Av., South Bend, IN
46634}

\vskip .2truein

\centerline{\it $^2$Kennesaw State University} \vskip .05truein
\centerline{\it 1000 Chastain Rd, Kennesaw, GA 30144}

\vskip .7truein

\baselineskip=19pt

\begin{quote}
 \centerline {\sc Abstract:}

 It is shown how the arithmetic structure of algebraic curves
 encoded in the Hasse-Weil L-function can be related to affine
 Kac-Moody algebras. This result is useful in relating the arithmetic
 geometry of Calabi-Yau varieties to the underlying exactly solvable theory.
 In the case of the genus three Fermat curve we identify the Hasse-Weil
 L-function with the Mellin transform of the twist of a number theoretic
 modular form derived from the string function of a non-twisted
 affine Lie algebra. The twist character is associated to the number
 field of quantum dimensions of the underlying conformal field theory.
 \end{quote}

\vfill

{\sc PACS Numbers and Keywords:} \hfill \break Math:  11G25
Varieties over finite fields; 11G40 L-functions; 14G10  Zeta
functions; 14G40 Arithmetic Varieties \hfill \break Phys: 11.25.-w
Fundamental strings; 11.25.Hf Conformal Field Theory; 11.25.Mj
Compactification

\renewcommand\thepage{}
\newpage



\parskip=-.02truein

\tableofcontents

\baselineskip=21pt

\parskip=.21truein
\parindent=0pt
\pagenumbering{arabic}

\section{Introduction}

{\bf 1.1} Number theoretic methods have proven useful in attempts
to understand string theoretic aspects of Calabi-Yau varieties.
Physically, string theory is a two-dimensional conformal field
theory on a Riemannian surface. Mathematically it can be viewed,
in the present context, as a vertex operator algebra associated to
an affine Kac-Moody algebra. The problem of string
compactification can be interpreted as an attempt to construct a
map which relates Kac-Moody vertex algebras to the geometry of
Calabi-Yau varieties. Part of the data of the conformal field
theory are the anomalous dimensions of the fields of the theory.
These scaling dimensions are rational numbers that appear in the
correlation functions of the quantum field theory. It turns out
that it is more useful to think about these values in terms of the
number field generated by their associated quantum dimensions,
defined via the characters of the Kac-Moody algebra. A relation
between these quantum dimensions and the arithmetic of Calabi-Yau
varieties of Brieskorn-Pham type has been established in
\cite{s03} by considering the Hecke theoretic nature of the
Hasse-Weil L-functions of these varieties. A more conceptual
framework of this relation has been formulated in \cite{lss04}.

An important aspect of conformal field theory is modular
invariance, a property of the string that appears difficult to
explain from a geometric point of view. It has in particular been
an open question for a long time how the string theoretic building
blocks on the world sheet are reflected in the geometry of
spacetime. One way to make this question more precise is by asking
how the characters that appear in the partition function of the
string emerge from spacetime, and how in turn the spacetime
geometry can be constructed from string theoretic quantities.
Questions of this type have a long history in arithmetic algebraic
geometry, e.g. in the context of the Shimura-Taniyama conjecture
\cite{w95, bcdt01} and the Langlands' program, but string theory
provides a somewhat different focus than the one encountered in a
broader arithmetic framework. Modularity of motivic L-functions
associated to algebraic varieties along the lines of Langlands'
program is not sufficient. In order to be of string theoretic
significance, the motivic modular forms should allow an
interpretation in terms of modular forms that arise in the
description of the physics on the Riemann surfaces defined by the
propagating string.

It was shown in \cite{su03} in the context of string
compactifications on elliptic curves that the Mellin transform of
the Hasse-Weil L-function of the plane cubic \beq C_3 = \{(x:y:z)
\in \mathP_2~|~x^3+y^3+z^3=0\} \eeq factors into a product of
modular forms that arise from the characters of the underlying
two-dimensional theory. More precisely, the following result was
obtained. Define $q=e^{2\pi i \tau}$ and let \beq \eta(\tau) =
q^{1/24} \prod_{n=1}^{\infty}(1-q^n)\eeq denote the Dedekind
$\eta$-function, $c^k_{\ell,m}(\tau)$ be the affine SU(2) string
functions at conformal level $k \in \mathN$, and
 \beq
 \Theta^k_{\ell,m}(\tau) = \eta^3(\tau) c^k_{\ell,m}(\tau)
 \lleq{su2theta}
be the Hecke indefinite modular forms associated to
$c^k_{\ell,m}$.

{\bf Theorem 1.1.} {\it The Mellin transform of the Hasse-Weil
L-function $L_{\rmHW}(C_3,s)$ of the cubic elliptic curve $C_3
\subset \mathP_2$ is a modular form $f_{\rmHW}(C_3,q) \in
S_2(\Gamma_0(27))$ which factors into the product} \beq
f_{\rmHW}(C_3,q) = \Theta(q^3)\Theta(q^9).\eeq {\it Here
$\Theta(\tau)=\eta^3(\tau)c^1_{1,1}(\tau)$ is the Hecke modular
form associated to the quadratic extension ${\mathbb Q}(\sqrt{3})$
of the rational field $\mathbb{Q}$, determined by the unique
string function $c^1_{1,1}(\tau)$ of the affine Kac-Moody
SU(2)-algebra at conformal level $k=1$.}

{\bf 1.2} Elliptic curves are somewhat degenerate examples of
Calabi-Yau manifolds and the question arises whether the results
of \cite{su03} can be extended to more general algebraic curves,
e.g. \beq C_n = \{(x:y:z)\in \mathP_2~|~x^n+y^n+z^n =0\}.
\lleq{planecurves} The genus of these Riemann surfaces is given by
$g(C_n)=(n-1)(n-2)/2$, hence except for the cubic curve $C_3$ just
discussed these curves are not elliptic. It is therefore not
obvious whether one should be able to find a string theoretic
interpretation, generalizing the result described above for $C_3$.
For such general Riemann surfaces there exists in particular no a
priori statement like the proof of the Shimura-Taniyama conjecture
which would ensure the modularity of the $q-$expansion derived
from the Hasse-Weil L-function. It is this question which we
address in this paper.

An argument in favor of a conformal field theoretic interpretation
of the curves $C_n$ is provided by the fact that they appear as
singular sets of higher dimensional Calabi-Yau varieties that are
expected to be exactly solvable. For the cubic curve $C_3$ such
higher dimensional manifolds are abundant, examples being provided
by the elliptic K3 surface defined by the degree 6 polynomial in
weighted projective space $\mathP_{(1,1,2,2)}$ \beq S_6 =
\{(z_1:\cdots :z_4)\in
\mathP_{(1,1,2,2)}~|~z_1^6+z_2^6+z_3^3+z_4^3=0\}, \eeq  or the
degree twelve Calabi-Yau threefold \beq X_{12} =
\{(z_0:\cdots:z_4)\in \mathP_{(1,1,2,4,4)}~|~z_0^{12}+z_1^{12} +
z_2^6 + z_3^3+z_4^3=0\}. \eeq To be more precise, the Gepner model
of $X_{12}$ is given by \beq \mathP_{(1,1,2,4,4)}\supset X_{12}
~\cong~ \left(10_A^2\otimes 4_A \otimes 1_A^2\right)_{\rm GSO},
\eeq with two minimal models at conformal level $k=10$, one model
at $k=4$, and two models at $k=1$, all equipped with the diagonal
affine invariant. The singular curve in this threefold is
$\mathP_{(1,2,2)}[6] \cong \mathP_2[3]$, i.e. the Fermat curve
$C_3$. The result of Theorem 1.1 provides a geometric construction
of the characters of the two minimal factors at $k=1$.

In the case of $X_{12}$ we can also identify the curve $C_3$ as
the generic elliptic fiber which can be identified by an iterative
application of the construction described in \cite{hs99}. First we
can construct the elliptic K3 surface $S_6$ via the twist map as
\beq \mathP_{(2,1,1)}[6] \times \mathP_2[3] \lra
\mathP_{(1,1,2,2)}[6], \eeq and then the hypersurface $X_{12}$ can
be constructed by applying the twist map again in the form \beq
\mathP_{(2,1,1)}[12] \times \mathP_{(1,1,2,2)}[6] \lra
\mathP_{(1,1,2,4,4)}[12]. \eeq We therefore see that the twist map
allows to construct the threefold $X_{12}$ as an iterated
orbifolds of products of curves and makes explicit the origin of
the two minimal factors at $k=1$. There is a large number of
varieties of this type among the Calabi-Yau hypersurface
threefolds constructed in the mirror paper \cite{cls90}, and in
refs. \cite{lgall}, which describe the complete construction of
this class of varieties.

For higher genus curves of type $C_n$ similar embeddings can be
obtained. A few examples are provided by the quartic plane curve
$C_4$, which is embedded in the octic hypersurface in weighted
projective space $\mathP_{(1,1,2,2,2)}$, the quintic curve $C_5$,
embedded in $\mathP_{(1,3,2,2,2)}$, or the septic curve $C_7$ in $
\mathP_{(1,7,2,2,2)}$. The class of higher dimensional varieties
of this type is large. Because they form part of an exactly
solvable variety we might expect these curves to inherit the
information of the underlying string theory. In this way we are
led to ask whether more general, non-Calabi-Yau varieties are
related to affine Kac-Moody algebras similar to the cubic.

In the present paper we first analyze the simplest non$-$elliptic
Fermat curve $C_4$ in detail, and in the last section indicate a
procedure which puts the quartic into perspective and shows that
other curves can be treated in a similar way. In particular we
prove the following result.

{\bf Theorem 1.2} {\it The Hasse-Weil L-function
$L_{\rmHW}(C_4,s)$ of the quartic plane curve $C_4$ factors into a
triple product $L_{\rmHW}(C_4,s) = L(E_4,s)^3$, where $L(E,s)$ is
Hasse-Weil L-functions of the weighted elliptic curve } \beq
E_4=\{(x_0:x_1:x_2)\in \mathP_{(1,1,2)} ~|~
x_0^4+x_1^4+x_2^2=0\}.\eeq {\it The Hasse-Weil modular form
$f(E_4,q)$ associated to $L(E_4,s)$ factors into a twisted product
\beq f(E_4,q)= \Theta^2_{1,1}(4\tau)^2\otimes
\chi_2.\lleq{quartic-modularity} Here the twist character is the
Legendre symbol $\chi_2(\cdot) = \left(\frac{2}{\cdot}\right)$ and
$\Theta^2_{1,1}(\tau)$ is the affine SU(2) theta function of the
string function $c^2_{1,1}(\tau)$ at conformal level $k=2$.}

This result shows that the modular forms derived from the quartic
curve admits a Kac-Moody theoretic interpretation via an SU(2)
theta function.

{\bf 1.3} We can turn this result around and read it as an
arithmetic
interpretation of a conformal field theoretic object, the string
theoretic parafermionic string functions. For any modular form
$f=\sum_n a_n q^n$ denote by $f_{\chi}(q) = \sum_n \chi(n)a_n q^n$
the form obtained by twisting $f$ by a character $\chi$.

{\bf Corollary 1.3} {\it The SU(2) string function
$c^2_{1,1}(\tau)$ at level $k=2$ is determined by the twisted
Hasse-Weil modular form} \beq c^2_{1,1}(\tau) =
\frac{1}{\eta^3(\tau)} \sqrt{f(E)\otimes \chi_2(q^{1/4})}.\eeq

{\bf 1.4} The weighted elliptic curve $E_4$ which emerges as the
arithmetic building block of the quartic curve also appears as the
singular set of many higher dimensional varieties and it is
possible to repeat the conformal field theoretic analysis
indicated above for Fermat curves. This can be illustrated by
considering the Calabi-Yau threefold $X_{16}$ of degree 16
embedded in the weighted projective space $\mathP_{(1,1,2,4,8)}$.
This variety is a K3-fibration with a generic fiber described by a
degree 8 hypersurface $S_8$ embedded in $\mathP_{(1,1,2,4)}$ which
is elliptic. The generic fiber of this elliptic fibration is
precisely the curve $E_4$ and we can iteratively construct the
degree sixteen threefold in $\mathP_{(1,1,2,4,8)}$ with the
methods of \cite{hs99} by first considering the map \beq
\mathP_{(2,1,1)}[8] \times \mathP_{(1,1,2)}[4] \lra
\mathP_{(1,1,2,4)}[8], \eeq and then applying the map \beq
\mathP_{(2,1,1)}[16] \times \mathP_{(1,1,2,4)}[8] \lra
\mathP_{(1,1,2,4,8)}[16]. \eeq Alternatively, $E_4$ appears as the
$\mathZ_2-$singular curve on this threefold. As in the case of
Theorem 1.1 in the case of the cubic curve Theorem 1.2 provides a
geometric interpretation of the string theoretic modular form of
the minimal factor at conformal level $k=2$ which is part of the
Gepner model corresponding to this threefold.

It becomes clear by an iterative application of the twist map
construction \cite{hs99} that elliptic threefolds which are K3
fibrations can be built from the generic elliptic fibre and other
curves. Thereby the question of modularity for these higher
dimensional varieties is reduced to the modularity problem of
curves.

{\bf 1.5} This paper is organized as follows. In Section 2 we
compute the Hasse-Weil L-function of the quartic. We will see that
arithmetically the basic building block of this curve is given by
the elliptic Brieskorn-Pham curve of degree four. In Section 2 we
briefly review the relevant aspects of non-twisted affine
Kac-Moody algebras. In Section 4 the modular form defined by the
twisted Mellin transform of the Hasse-Weil L-function is related
to a modular form derived from the character of an affine
Kac-Moody algebra. A relation between the character defining this
twist and the quantum dimensions of the affine Kac-Moody algebra
is described in Section 5. In the final Section 6 we show that the
factorization behavior of the quartic is typical for Fermat
curves, indicating that other curves can be treated in a similar
way.

\vskip .3truein

\section{Hasse-Weil L-Function}

{\bf 2.1} For algebraic varieties $X$ the congruent zeta function
at a prime number $p$ can be defined as the generating function
\beq Z(\XIFp, t) \equiv exp\left(\sum_{r\in \IN} \#
\left(\XIFpr\right) \frac{t^r}{r}\right).\eeq It was first shown
by F.K. Schmidt \cite{fks31, h34} that for algebraic curves $X$
the zeta function $Z(\XIFp,t)$ is a rational function which takes
the form \beq Z(\XIFp,t) = \frac{\cP^{(p)}(t)}{(1-t)(1-pt)},\eeq
where \beq \cP^{(p)}(t)= \sum_{i=0}^{2g} \beta_i(p)t^i, \eeq is a
polynomial whose degree is given by the genus $g(X)=(2-\chi(X))/2$
of the curve, where $\chi(X)$ is the Euler characteristic of the
curve. More important is the global zeta function obtained by
setting $t=p^{-s}$ and taking the product over all rational primes
at which the variety has good reduction. Let $S$ denote the set of
rational primes at which $X$ becomes singular and denote by $P_S$
the set of primes that are not in $S$. The global zeta function
then becomes \beq Z(X,s) = \prod_{p \in P_S}
\frac{\cP^{(p)}(p^{-s})}{(1-p^{-s})(1-p^{1-s})} =
\frac{\zeta(s)\zeta(s-1)}{L_{\rmHW}(X,s)},\eeq with the Hasse-Weil
L-function \beq L_{\rmHW}(X,s) = \prod_{p \in P_S}
\frac{1}{\cP^{(p)}(p^{-s})},\lleq{lhw} and the Riemann zeta
function $\zeta(s) = \prod_p (1-p^{-s})^{-1}$.

The most direct way to compute the expansion at least for the
lower primes involves the comparison of definition of the
congruent zeta function with the rational expression found by
Schmidt. For our purposes it suffices to collect
the first three nontrivial coefficients \bea \beta_0(p) &=& 1 \nn \\
\beta_1(p) &=& N_{1,p} -(p+1) \nn \\  \beta_2(p) &=&
\frac{1}{2}(N_{1,p}^2 + N_{2,p}) - (p+1)N_{1,p} + p \nn \\
\beta_3(p) &=& \frac{1}{3}N_{3,p} + \frac{1}{2}N_{2,p} +
\frac{1}{6}N_{1,p}^3 - \frac{1}{2}(N_{1,p}^2 + N_{2,p}) + pN_{1,p}
\nn \\  &\vdots & \nn \\ \beta_6(p) &=& p^3.\eea We collect in
Table 1 our results for the first few primes.
\begin{center}
\begin{tabular}{l | c c c c c c c c c c c c c }
Prime $p$     &2  &3   &5    &7  &11  &13  &17  &19  &23
              &29 &31  &37   &41  \tabroom \\
\hline
$N_{1,p}$     &3  &4   &0    &8  &12  &32  &12  &20  &24
              &0  &31  &32   &12   \tabroom \\
$N_{2,p}$     &5  &28  &34   &   &    &    &    &    &
              &   &    &     &    \tabroom \\
\hline
$\beta_1(p)$  &0  &0   &$-6$ &0  &0   &18  &$-6$ &0   &0
              &$-30$ &0 &$-6$ &$-30$  \tabroom \\
$\beta_2(p)$  &0  &9   &27   &   &    &    &     &    &
              &   &    &     &   \tabroom \\
\hline
\end{tabular}

{\bf Table 1.}~{\it The coefficients $\beta_1,\beta_2$ for the
quartic curve $C_4$ of genus 3.}
\end{center}

Using these results leads to the expansion \bea L_{\rmHW}(C_4,s)
&=& \prod_p \frac{1}{1+\beta_1(p)p^{-s} + \cdots + p^3 p^{-6s}}
\nn \\ &=& 1 + \frac{6}{5^s} - \frac{9}{9^s} - \frac{18}{13^s} +
\frac{6}{17^s} + \frac{9}{25^s} +\frac{30}{29^s} + \frac{6}{37^s}
+ \frac{30}{41^s} + \cdots \eea The associated $q-$series then
takes the form \beq f_{\rmHW}(C_4,q) = q + 6q^5 - 9q^9 - 18q^{13}
+ 6q^{17} + 9q^{25} + 30q^{29}+ 6q^{37} + 30q^{41} + \cdots
\lleq{hwmodform} Finite expansions like this often turn out to be
useful because of theorems by Faltings and Serre which show that
such functions are determined uniquely by a finite number of
terms.

{\bf 2.2} The behavior of the coefficients $a_n$ of the Hasse-Weil
$q-$expansion under Hecke operators indicates that this series
does not describe a Hecke eigenform. Hence the above result for
the Hasse-Weil L-function is not particularly illuminating except
for the fact that all the coefficients are divisible by 3. This
suggests that the Hasse-Weil L-function of the quartic plane curve
can be viewed as the cubic power of a more basic L-series. We can
write $L_{\rmHW}(C_4,s) = L^3(s)$ with \beq L(s) = 1 +
\frac{2}{5^s} - \frac{3}{9^s} - \frac{6}{13^s} + \frac{2}{17^s} -
\frac{1}{25^s} +\frac{10}{29^s} + \frac{2}{37^s} + \frac{10}{41^s}
+ \cdots \eeq The series $L(s)$ can be obtained via the Mellin
transform \beq L(s) = \frac{(2\pi)^s}{\Gamma(s)} \int_0^{\infty}
f(iy) y^{s-1}dy \eeq from the expansion \beq f(q) = q + 2q^5 -
3q^9 - 6q^{13} + 2q^{17} - q^{25} + 10q^{29}+ 2q^{37} + 10q^{41} +
\cdots \lleq{cuberoot} This result indicates that $f(q)$ is the
basic building block of the Hasse-Weil L-form $f_{\rmHW}(C_4,q)$,
providing a sort of cubic root of it via its L-series.

{\bf 2.3} The question arises whether one can interpret the
expansion (\ref{cuberoot}) in a geometric way as the Hasse-Weil
modular form of some other geometric object. To answer such a
factorization question it is useful to understand the polynomials
$\cP^{(p)}(t)$ which determine the congruence zeta function in a
more systematic way in terms of Jacobi sums.

{\bf Theorem.} \cite{w52} {\it For the plane curve} \beq C_n
=\{z_0^n +z_1^n +z_2^n =0 \} \subset \mathP_2 \eeq {\it defined
over the finite field $\IF_q$ set $d=(n,q-1)$ and define the set
$\cA_2^{p,n}$ of triplets $\a =(\a_0,\a_1,\a_2)$
 of rational numbers } \beq \cA_2^{q,n} = \left\{
(\a_0,\a_1,\a_2) \in {\mathbb Q}^3~|~0<\a_i<1,~d=(n,q-1),~d\a_i
=0~(\rmmod~1),~\sum_{i=0}^2 \a_i=0~(\rmmod~1)\right\}.\eeq {\it
For such triplets define the Jacobi sums } \beq
j_q(\a_0,\a_1,\a_2) = \frac{1}{q-1} \sum_{\stackrel{u_0,u_1,u_2
\in \IF_q}{u_0+u_1+u_2=0}} \chi_{\a_0}(u_0) \chi_{\a_1}(u_1)
\chi_{\a_2}(u_2) \eeq {\it where $\chi_{\a_i}(u_i) = e^{2\pi i
\a_i m_i}$ and the integers $m_i$ are determined via $u_i =
g^{m_i}$, where $g\in \IF_q$ is a generator. Then the cardinality
of $C_n/\IFq$ is given by} \beq \#(C_n/\IF_q) = N_{1,q}(C_n) = 1+q
+ \sum_{(\a_0,\a_1,\a_2)\in \cA_2^p} j_q(\a_0,\a_1,\a_2).\eeq

With these Jacobi sums one can express the F.K. Schmidt polynomial
as \beq \cP^{(p)}(t) = \prod_{\a \in
\cA_2^n}\left(1+j_{p^{\mu_{\a}}}(\a)t^{\mu_{\a}}\right)^{1/\mu_{\a}},
\eeq where $\mu_{\a}$ is the smallest positive integer such that
for $\a$ in the set $\cA_2^n$ defined by \beq \cA_2^n = \left\{\a
\in \mathQ^3~|~ \a_i \in (0,1), n\a_i \in \mathN, \sum_{i=0}^2
\a_i \in \mathN \right\}, \eeq one finds $(p^{\mu_{\a}}-1)\a_i \in
\mathN$ for all $i$. Applied to the quartic curve $C_4$ the set
$\cA_2^{q,4}$ is given by \beq \cA_2^{q,4} =
\left\{\begin{tabular}{c l}
        $\emptyset$,  &if $d=(4,q-1)\in \{1,2\}$ \\
        $\cA_2^4$     &if $d=(4,q-1) =4$. \\
   \end{tabular}
   \right\}
   \eeq
   with $\cA_2^4 = \cA_2^4(1)\cup \cA_2^4(2)$ where  \bea
\cA_2^4(1) &=& \left\{\footnotesize
\left(\frac{1}{2},\frac{1}{4},\frac{1}{4}\right),
\left(\frac{1}{4},\frac{1}{2},\frac{1}{4}\right),
\left(\frac{1}{4},\frac{1}{4},\frac{1}{2}\right)\right\} \nn \\
\cA_2^4(2) &=& \left\{\footnotesize
\left(\frac{1}{2},\frac{3}{4},\frac{3}{4}\right),
\left(\frac{3}{4},\frac{1}{2},\frac{3}{4}\right),
\left(\frac{3}{4},\frac{3}{4},\frac{1}{2}\right)\right\}, \eea The
sums $j_q(\a)$ are constant over the two permutation orbits of
$\cA_2^4$, and with $j_p\left({\footnotesize \frac{1}{2},
\frac{3}{4}, \frac{3}{4}}\right)=
\overline{j_p\left({\footnotesize
\frac{1}{2},\frac{1}{4},\frac{1}{4}}\right)}$ one finds \beq
L_{\rmHW}(C_4,s) = \prod_p \frac{1}{\left(1+2a_p\cdot p^{-s} +
p\cdot p^{-2s}\right)^3},\eeq where $a_p=\rmRe
~j_p\left({\footnotesize
\frac{1}{2},\frac{1}{4},\frac{1}{4}}\right)$.

Hence $L_{\rmHW}(C_4,s) = L^3(s)$ in terms of an L-function \beq
L(s) = \prod_p \frac{1}{1+2a_p\cdot p^{-s} + p\cdot
p^{-2s}}\lleq{ellhw} which has the form of the Hasse-Weil series
of an elliptic curve. The quartic curve $C_4$ has genus $g=3$,
hence this factorization indicates that the L-function of this
curve splits into a product of $g$ elliptic contributions. Such a
factorization into elliptic factors will not happen for general
Fermat curves.

{\bf 2.4} The above decomposition allows us to return to the
question asked above whether one can determine an elliptic curve
whose Hasse-Weil series has been determined by (\ref{ellhw}). The
quartic curve $C_4$ becomes singular when reduced at $p=2$. Hence
we would expect the purported elliptic curve to have a conductor
that is divisible by two, suggesting an elliptic curve of degree
four. Such a curve is the well known weighted torus, described by
\beq E=\{(x_0:x_1:x_2)\in \mathP_{(1,1,2)} ~|~
x_0^4+x_1^4+x_2^2=0\}.\eeq The cardinalities for this curve are
collected in Table 3.


\begin{center}
\begin{tabular}{l| r r r r r r r r r r r r r}
Prime $p$ &2 &3 &5     &7  &11  &13   &17     &19   &23  &29    &31  &37  &41\tabroom \\
\hline
$N_{1,p}$ &3 &4 &4     &8  &12  &20   &16     &20   &24  &20    &32  &36  &32\tabroom \\
\hline
$\b_1(p)$ &0 &0 &$-2$  &0  &0   &6    &$-2$   &0    &0   &$-10$ &0  &$-2$  &$-10$\tabroom \\
\hline
\end{tabular}
\end{center}
{\bf Table 2.}{\it ~~Coefficients $\b_1(p)=N_{1,p}(E)-(p+1)$ of
the Hasse-Weil modular form of the elliptic quartic curve $E$ in
terms of the cardinalities $N_{1,p}(E)$ for the lower rational
primes.}

The Hasse-Weil modular form resulting from the solutions $E/\IF_p$
starts out as \beq f_{\rmHW}(E,q) = q +2q^5 -3q^9 -6q^{13}
+2q^{17} -q^{25} +10q^{29} +2q^{37} + 10q^{41} + \cdots
\lleq{basicL} in agreement with the expansion $f(q)$ associated to
the 'cube root' $L(s)$ discussed above.

We are therefore in the same situation as in \cite{su03} and we
can ask whether the modular form $f(q)=f_{\rmHW}(E,q)$ can be
related to modular forms that are induced by conformal field
theoretic characters.

\vskip .3truein

\section{Affine Kac-Moody Algebras}

{\bf 3.1} A construction of non-twisted affine Kac-Moody algebras,
also called affine Lie algebras, is provided by the extension
\cite{k90} \beq \hurmG = \rmL\urmG \oplus \mathC k \oplus \mathC d
\eeq of the loop algebra \beq \rmL\urmG = \urmG \otimes
\mathC[t,t^{-1}]\eeq by the central extension $k$ and
$d=t\frac{d}{dt}$. In terms of the generators $J^a\otimes t^m$ the
algebra becomes \beq [J^a\otimes t^m,J^b\otimes t^n] =
if^{ab}_cJ^c \otimes t^{m+n} + km \delta^{ab} \delta_{m+n,0}. \eeq
The representations of this algebra can be parametrized by affine
roots $\hat{\lambda}=(\lambda,k,n)$ of the Cartan-Weyl subalgebra
$\{H^i_0,E^{\alpha}_0,L_0\}$, with $i=1,...,r$, where $r$ denotes
the rank of the underlying Lie algebra $\urmG$. For fixed level
$k$ of the theory the characters $\chi_{\hat{\lambda}}$ are
essentially parametrized by the weight $\lambda$ of the
representation.  For the reduced characters of an affine Lie
algebra $\uhG$ at level $k$ the characters transform as \beq
\chi_{\hat{\lambda}}(-1/\tau) =
       \sum_{\hat{\mu} \in P_+^k} S_{\hat{\lambda},\hat{\mu}}
       \chi_{\hat{\mu}}(\tau),\eeq where the modular $S$-matrix takes
       the form  \beq S_{\hat{\lambda},\hat{\mu}} =
        \frac{i^{|\Delta_+|}}{\sqrt{|P/Q^{\vee}| (k+g)^r}} \sum_{w\in W}
        \epsilon(w) e^{-2\pi
        i\frac{<w(\lambda+\rho),\mu+\rho>}{k+g}}.\eeq
        Here $P=\sum_i {\mathZ} \omega_i$ denotes the lattice
        with fundamental weights $\omega_i$ defined by
        $<\omega_i,\alpha_j^{\vee}>=\delta_{ij}$ via co-roots
        $\alpha_j^{\vee}=\frac{2\alpha}{<\alpha,\alpha>}$.
        $Q^{\vee}=\sum_i {\mathZ}\alpha_i^{\vee}$ is the coroot lattice
        and $P/Q^{\vee}$ denotes the lattice points of $P$ lying in an elementary
         cell of $Q^{\vee}$, while $|P/Q^{\vee}|$ describes the number of points in this
         set. $P_+^k$ is the set of all dominant weights at level
         $k$, and $\epsilon(w)=(-1)^{\ell(w)}$ is the signature of the
         Weyl group element $w$,
         where $\ell(w)$ is the minimum number of simple Weyl reflections that $w$
         decomposes into. $\Delta_{+}$ is the number of positive roots in
         the Lie algebra $\underline{G}$.

{\bf 3.2} Supersymmetric string models can be constructed in terms
of conformal field theories with N=2 supersymmetry. The simplest
class of N=2 supersymmetric exactly solvable theories is built in
terms of the affine SU(2)$_k$ algebra at level $k$ as a coset
model ${\rm SU(2)}_k \otimes {\rm U(1)}_2/{\rm U(1)}_{k+2,{\rm
diag}}$. Coset theories $G/H$ lead to central charges of the form
$c_G - c_H$, hence the supersymmetric affine theory at level $k$
still has central charge $c_k=3k/(k+2)$. The spectrum of anomalous
dimensions $\Delta^{\ell}_{q,s}$ and U(1)$-$charges $Q^{\ell}$ of
the primary fields $\Phi^{\ell}_{q,s}$ at level $k$ is given by
\begin{eqnarray} \Delta^{\ell}_{q,s} &=&
\frac{\ell (\ell +2)-q^2}{4(k+2)} + \frac{s^2}{8} \nonumber \\
Q^{\ell}_{q,s} &=& - \frac{q}{k+2} + \frac{s}{2}, \end{eqnarray}
where $\ell\in \{0,1,\dots,k\}$, $\ell+q+s \in 2\mathbb{Z}$, and
$|q-s|\leq \ell$. Associated to the primary fields are characters
defined as
\begin{equation}
\chi^k_{\ell,q,s}(\tau, z,u) = e^{-2\pi i u} {\rm
tr}_{\cH^{\ell}_{q,s}} e^{2\pi i\tau (L_0 -\frac{c}{24})} e^{2\pi
i J_0}, \end{equation} where the trace is to be taken over a
projection $\cH^{\ell}_{q,s}$ to a definite fermion number (mod 2)
of a highest weight representation of the (right-moving) $N=2$
algebra with highest weight vector determined by the primary
field. It is of advantage to express these maps in terms of the
string functions and theta functions, leading to the form
\begin{equation} \chi^k_{\ell,q,s}(\tau, z, u) = \sum
c^k_{\ell,q+4j-s}(\tau) \theta_{2q+(4j-s)(k+2),2k(k+2)}(\tau,
z,u).
\end{equation} It follows from this representation that
the modular behavior of the $N=2$ characters decomposes into a
product of the affine SU(2) structure in the $\ell$ index and into
$\Theta$-function behavior in the charge and sector index. The
string functions $c^k_{\ell,m}$ are given by \beq
c^k_{\ell,m}(\tau) = \frac{1}{\eta^3(\tau)}
\sum_{\stackrel{\stackrel{-|x|<y\leq |x|}{(x,y)~{\rm
or}~(\frac{1}{2}-x,\frac{1}{2}+y)}}{\in
\mathZ^2+\left(\frac{\ell+1}{2(k+2)},\frac{m}{2k}\right)}}
\rmsign(x) e^{2\pi i \tau((k+2)x^2-ky^2)} \lleq{strngfncts}
 while the classical
theta functions $\theta_{m,k}$ are defined as  \beq
\theta_{n,m}(\tau,z,u) = e^{-2\pi i m u} \sum_{\ell \in \mathZ +
\frac{n}{2m}} e^{2\pi i m \ell^2 \tau + 2\pi i \ell z}.\eeq It
follows from the coset construction that the essential ingredient
in the conformal field theory is the SU(2) affine theory.

{\bf 3.3} It was suggested by Gepner some time ago that exactly
solvable string compactifications obtained by tensoring several
copies of $N=2$ supersymmetric conformal field theories derived
from affine Kac-Moody algebras should yield, after performing
appropriate projections, theories that correspond in some limit to
geometric compactification described by Brieskorn-Pham type
Calabi-Yau varieties \cite{g88}. The evidence for this conjecture
was initially based mostly on spectral information for all models
in the Gepner class of solvable string compactifications and the
agreement of certain types of intersection numbers which allow an
interpretation as Yukawa couplings, as well as Landau-Ginzburg
type arguments \cite{mlvw89}. In the case of the Fermat cubic
curve these results suggest that there is an underlying conformal
field
 theory of this elliptic curve that is described by the GSO
 projection of
 a tensor product of three models at conformal level $k=1$.
Roughly, this entails a relation of the type \beq \mathP_2 \supset
C_3 ~~\lolra ~~\left(\rmSU(2)_{k=1,A_1}\right)^{\otimes 3}_{\rm
GSO},\lleq{exactcubic} where $A_1$ signifies the diagonal
invariant for the SU(2) partition function, and GSO indicates the
projection which guarantees integral U(1)-charges of the states.

In the case of higher genus curves no such direct relation is
expected from a string theory perspective. There is, however, a
weaker embedding argument that suggests a possible modular
interpretation. Higher genus curves can be embedded in higher
dimensional Calabi-Yau varieties which in turn are conjectured to
be exactly solvable. Examples of such embeddings for the quartic
curve are provided by threefolds such as Brieskorn-Pham
hypersurfaces $\mathP_{(1,1,2,2,2)}[8]$,
$\mathP_{(1,2,3,3,3)}[12]$, $\mathP_{(1,4,5,5,5)}[20]$, and
others. The first of these varieties e.g. \beq X_8 = \{z_0^8+z_1^8
+ z_2^4 + z_3^4 + z_4^4=0\} \subset \mathP_{(1,1,2,2,2)} \eeq is
expected to be related to the conformal field theory \beq
\left(\left(\rmSU(2)_{k=6,A_1}\right)^{\otimes 2} \otimes
\left(\rmSU(2)_{k=2,A_1}\right)^{\otimes 3}\right)_{\rm GSO}.\eeq
Hence we might expect that conformal field theoretic aspects are
encoded in the quartic curve  \beq \mathP_2 \supset C_4 \lolra
\left(\rmSU(2)_{k=2,A_1}\right)^{\otimes 3}.\lleq{exactquartic}

This leads to the affine Kac-Moody algebra SU(2) at level two and
we can ask whether there are relations between the modular objects
determined by the affine algebra, and the modular objects
determined by the variety. The results of \cite{su03} suggest that
interesting affine quantities to consider are the SU(2) theta
functions associated to string functions at conformal level $k$,
defined as $\Theta^k_{\ell,m}(\tau) = \eta^3(\tau)c^k_{\ell,m}$.
The expansion of the theta functions at conformal level $k=2$
follows from those of the string function expansions. It turns out
that relevant for the present discussion is the theta function
\beq \Theta^2_{1,1}(q) = q^{1/8} (1 - q - 2q^2 + q^3 + 2q^5 +
\cO(q^6)). \eeq

This is a modular form of weight one, and therefore cannot be
identified directly with the Hasse-Weil form itself. It will
become clear below, however, that $\Theta^2_{1,1}(q)$ emerges  as
the building block of the Hasse-Weil modular form
$f_{\rmHW}(C_4,q)$ of the quartic plane Fermat curve.

\vskip .3truein

\section{Geometric Modularity}

{\bf 4.1} From a physical perspective it is not clear a priori
which conformal field theoretic quantities should be the correct
building blocks of the Hasse-Weil function, if any. Possibilities
include the twists of the affine or parafermionic characters, or
some elements of the $N=2$ superconformal model. The coset
construction shows that the basic part of the $N=2$ theory is
given by the affine SU(2) Kac-Moody algebra. The string functions
$c^k_{\ell,m}(\tau)$ of the $N=2$ characters would appear to be
natural candidates because they capture the essential interacting
nature of the field theory. Furthermore, associated to string
functions are natural number theoretic theta functions.

The main result of \cite{su03} shows that the Hasse-Weil
L-function of the cubic plane curve $C_3$ is determined by the
Kac-Moody string function of the affine SU(2) algebra at conformal
level $k=1$.  At $k=1$ there is only one string function, denoted
by $c(\tau)$, which can be computed to lead to the expansion \beq
c(\tau) = q^{-1/24}(1+q+2q^2 + 3q^3 + 5q^4 + 7q^5 + \cO(q^6)).
\eeq It turns out that more important than the string functions
are the associated SU(2) theta functions $\Theta^k_{\ell,m}(\tau)$
(\ref{su2theta}). These objects are associated to quadratic number
fields determined by the level of the affine theory \cite{kp84}.
At level $k=1$ the unique theta function $\Theta(\tau)$ is
associated to the real quadratic extension ${\mathbb Q}(\sqrt{3})$
of the rational field $\mathbb{Q}$. Its expansion follows from the
string function expansion, resulting in \beq \Theta(q) = q^{1/12}
(1 - 2q - q^2 + 2q^3 + q^4 + 2q^5 + \cO(q^6)). \eeq It is this
modular form of weight one which emerged in \cite{su03} as the
building block of the Hasse-Weil modular form $f_{\rmHW}(C_3,q)$
of the cubic elliptic curve \beq f_{\rmHW}(C_3,\tau) =
\Theta(3\tau)\Theta(9\tau).\eeq

{\bf 4.2} The case of the elliptic cubic curve is special because
it is a Calabi-Yau curve and therefore defines a consistent string
theory background. For higher genus curves this is not the case,
and it is therefore not a priori clear whether one should expect a
relation to the string theoretic Kac-Moody algebra. As mentioned
above, an argument that is encouraging is that higher genus
algebraic curves such as the quartic, and others, appear in higher
dimensional Calabi-Yau varieties as singular curves which have to
be resolved.

{\bf 4.3} In the case of the quartic curve we can use the result
derived above that the L-function is that of a triple product of
elliptic curves. This reduces the current problem to the type of
problem solved in \cite{su03}, and we can follow the logic used in
the construction introduced there. First we need to determine the
weight and the level of this form. For a general elliptic curve
the corresponding modular form $f_{\rmHW}(E,q)$ defined by the
Mellin transform of the Hasse-Weil L-function $L_{\rmHW}(E,s)$ is
determined by the proof of the Shimura-Taniyama conjecture to be
an element in $S_2(\Gamma_0(N))$ for some level $N$.
Alternatively, one can read off the weight of a Hecke eigenform
from the multiplicative properties of its coefficients
$\{a_n\}_{n\in
\mathN}$ induced by the Hecke operators
 \bea a_{mn}&=& a_ma_n~~~(m,n)=1 \nn \\
a_{p^{n+1}} &=& a_{p^n}a_p - p^{k-1}a_{p^{n-1}} \nn \\  a_{p^n}
&=& (a_p)^n,~~~~{\rm for}~p|N, \eea as described in more detail in
\cite{su03} for the case of cubic plane curve.

{\bf 4.4} The quartic curve is singular at the prime $p=2$,
therefore we expect the conductor of the elliptic curves involved
to be divisible by 2, and perhaps by powers of two. Using Weil's
conductor conjecture \cite{w67} for elliptic curves we further
expect the modular conductor of the corresponding modular form to
be divisible by some power of 2. The arithmetic conductor of the
elliptic curve $E$ can be determined via Tate's algorithm
\cite{t75} from the generalized affine form of $E$ given by \beq
y^2 - a_1xy - a_3y = x^3 + a_2x^2 + a_4x + a_6. \eeq By a result
of Ogg \cite{o67} the conductor of such a curve can be computed as
\beq N_{E/\mathbb{Q}} = \prod_{{\rm bad}~p} p^{f_p},\eeq where the
exponent $f_p$ is given by \beq f_p = \rmord_p
\Delta_{E/\mathbb{Q}} + 1 - s_p, \eeq in terms of discriminant
$\Delta_{E/\mathQ}$ of the curve and the number $s_p$ of
irreducible components of the singular fiber at $p$. This shows
that the conductor of the elliptic curve can be viewed as a
quantity which combines the bad primes with a measure of the
severity of the singularity at these bad primes.

{\bf 4.5} The elliptic quartic curve $E_4$ can be transformed into
its affine form by choosing inhomogeneous coordinates and defining
new coordinates $v=xz/y$ and $u=(x/y)^2$, leading to the
Weierstrass form \beq v^2 = u^3 + u.\eeq Tate's computation of the
discriminant simplifies considerably for this curve, leading to
$\Delta_{E_4}=-64$, and therefore $\rmord_2~\Delta_{E_4}=6$. The
singular fiber is of Kodaira type II, and therefore the arithmetic
conductor is $N=64$.

{\bf 4.6} Combining the conductor argument with the string
theoretic embedding of the quartic into higher dimensional
varieties we are led to consider the form \beq
\Theta^2_{1,1}(q^4)^2 = q-2q^5 - 3q^9+6q^{13} + 2q^{17} -q^{25}
-10q^{29} + \cdots \eeq Comparing this with the Mellin transform
(\ref{basicL}) of the cubic root of the Hasse-Weil L-form of the
quartic $C_4$ shows agreement with $\Theta^2_{1,1}(4\tau)^2$
except for the signs in the terms $q^n$ with $n=1~\rmmod~8$.

We therefore see that the discussion of the quartic curve must
involve an ingredient that goes beyond the analysis that succeeded
in \cite{su03} in providing a string theoretic Kac-Moody algebra
interpretation of the Hasse-Weil L-function. In some sense the
elliptic curve defined by $C_3$ is too special to reveal all the
key elements necessary for this identification. In the following
we will first complete the identification of the Hasse-Weil
modular form with a string theoretic modular form in a somewhat
utilitarian manner by pin-pointing the missing ingredient. We then
turn to the physical interpretation of this ingredient and explain
why it does not appear in the discussion of $C_3$.

The sign flip suggests that the Hasse-Weil modular form $f_{\rmHW}(E_4,q)$
 might be related to the modular form
$\Theta^2_{1,1}(4\tau)^2$ via a twist. For a modular form
$f(q)=\sum_n a_nq^n$ and a Dirichlet character \beq \chi:~\mathZ
\lra K^{\times} \eeq with values in a field $K$, define the
twisted from as \beq f_{\chi}(q) = \sum_n \chi(n)a_nq^n.\eeq We
are interested in a character with conductor 8, taking values in
$K=\IF_2$. An example of a class of characters which leads to such
an object is provided by Legendre symbols. These are defined on
rational primes as \beq \chi_n(p) = \left(\frac{n}{p}\right) =
\left\{\begin{tabular}{c l} ~1 &$n$ is a square in $\IF_p$ \\ $-1$
&$n$ is not a square in
$\IF_p$ \tabroom \\
\end{tabular} \right\} \eeq The conductor of $\chi_n(\cdot)$ is
given by $n$ if $n=1~(\rmmod~4)$ and $4n$ for $n=2,3~(\rmmod~4)$.

For non-prime numbers the generalized Legendre symbol is defined
by using the prime decomposition. Every natural number $m$ can be
decomposed into primes as $m=p_1\cdots p_r$ and the generalized
symbol is defined as \beq \chi_n(m) = \prod_{i=1}^r
\left(\frac{n}{p_i}\right).\eeq This shows that for $n=2$ the
Legendre symbol takes the form \beq \chi_2(r) =
\left(\frac{2}{r}\right) = \left\{
\begin{tabular}{c l}
~1 &$r=1~\rmmod~8$ \\
$-1$ &$r=5~\rmmod~8$ \tabroom \\
\end{tabular}\right\}  \eeq

It is this character which provides the twist from the conformal
field theory induced modular form to the elliptic  L-function
which provides the basic building block of the Hasse-Weil
L-function of the quartic curve.

We therefore see that the CFT modular form
$\Theta^2_{1,1}(4\tau)^2$ of weight two is the twist of the
geometric weight two modular form via the number theoretic
Legendre symbol. Put differently, we see that if we denote by
$\Theta^2_{1,1}(4\tau)^2\otimes \chi_2$ the modular form obtained
by twisting the form $\Theta^2_{1,1}(4\tau)^2$ by the character
$\chi_2(\cdot)$, we can write the Hasse-Weil L-function of as \beq
L_{\rmHW}(C_4,s) = (L(\Theta^2_{1,1}(4\tau)^2\otimes
\chi_2,s)^3.\eeq

This concludes the proof of the Theorem stated in the
Introduction.

The remaining question is whether the theta function input is
unique. To answer this one can use the Eichler-Shimura theory
\cite{e54, s58, s59}. $\Theta^2_{1,1}(4\tau)^2$ is an element in
the space of cusp forms $S_2(\Gamma_0(32))$. For arbitrary modular
level $N$, the dimension of the space $S_2(\Gamma_0(N))$ can be
computed as the genus of the modular curve $X_0(N)$ via  \beq
g(X_0(N)) = 1+ \frac{\mu(N)}{12} - \frac{\nu_2(N)}{4} -
\frac{\nu_3(N)}{3} - \frac{\nu_{\infty}}{2},\eeq where $\mu(N)$ is
the index of $\Gamma_0(N)$ in $\Gamma(1)=\rmSL(2,\mathZ)$,
$\nu_2(N)$ and $\nu_3(N)$ are the number of elliptic points of
order 2 and 3, and $\nu_{\infty}(N)$ is the number of
$\Gamma_0(N)$ inequivalent cusps. For $N=32$ this implies that the
space of cusp forms is one dimensional, hence
$\Theta^1_1(4\tau)^2$ is its unique generator, up to the
multiplication of a constant.

\vskip .3truein

\section{Quantum Dimensions}

The character $\chi_2(\cdot)$ which appears above in the context
of providing a string theoretic interpretation of the Hasse-Weil
modular form furthermore points toward the field of quantum
dimensions, thereby providing a link between the problem of a
geometric explanation of the string theoretic modularity and the
problem of providing a geometric explanation of the string
theoretic spectrum.

For general square free $n$ the Legendre characters
$\chi_n(\cdot)$ allow a characterization of the factorization
behavior of rational primes $p$ when viewed as elements in a
quadratic extension $\mathQ(\sqrt{n})$. They describe whether the
rational prime $p$ splits into a product of prime ideals $\pfrak_i
\subset \cO_{{\mathbb{Q}}(\sqrt{n})}$ in the ring of algebraic
integers $\cO_{\mathQ(\sqrt{n})}$. The principal ideal $(p)$
factors as $(p)=\pfrak_1 \pfrak_2$ if $\chi_n(p)=1$, remains prime
if $\chi_n(p)=-1$, and ramifies, i.e. $(p)=\pfrak^2$, if
$\chi_n(p)=0$, i.e. $p|n$. In summary we can write \beq \chi_n(p)
= \left\{ \begin{tabular}{r l}
                         1  &if $(p) = \pfrak_1\pfrak_2$ \\
                         $-1$ &if $(p)$ is prime \tabroom \\
                         0    &if $(p)=\pfrak^2$ \tabroom \\
         \end{tabular}
  \right\}.
\eeq For $n=2$ the character $\chi_2$ therefore is associated to
the quadratic extension $\mathQ(\sqrt{2})$. It turns out that this
is precisely the field determined by the anomalous dimensions of
the affine theory when mapped into the quantum dimensions. This
can be seen as follows.

It was noted in \cite{s03} that a link between the geometry and
the conformal field theory can be obtained by using a translation
of the anomalous dimensions into algebraic integers via the Rogers
dilogarithm. In the simple case of the affine Lie algebra SU(2) at
conformal level $k$ the modular transformations of the characters
$\chi$ associated to the primary fields $\Phi_i$ with anomalous
dimensions \beq \Delta_{\ell} = \frac{\ell(\ell+2)}{4(k+2} \eeq
are given by \beq \chi_{\ell}\left(-\frac{1}{\tau},
\frac{u}{\tau}\right) = e^{\pi i ku^2/2} \sum_m S_{\ell m}
\chi_m(\tau, u)\eeq with modular $S$-matrix
 \beq S_{\ell m} = \sqrt{\frac{2}{k+2}}
~~\sin\left(\frac{(\ell+1)(m+1)\pi}{k+2}\right),~~~~~0\leq \ell,m
\leq k. \eeq

With these matrices one can define the generalized quantum
dimensions as $Q_{\ell m} = \frac{S_{\ell m}}{S_{0m}}$ for the
affine SU(2) algebra at level $k$. The importance of these numbers
derives from the fact that even though they do not directly
provide the scaling behavior of the correlation functions, they do
contain the complete information about the anomalous dimensions as
well as the central charge. The first step in this direction was
the realization by Kirillov and Reshetikhin that the central
charge can be expressed in terms of the quantum dimensions.
Earlier mathematical results had been obtained by Lewin. Denote by
 $L$  Rogers' dilogarithm \beq L(z) = Li_2(z) +{\small
\frac{1}{2}} \log(z)~\log(1-z) \eeq and $Li_2$ is Euler's
classical dilogarithm \beq Li_2(z) = \sum_{n\in \IN}
\frac{z^n}{n^2}.\eeq Then one has the following result.

{\bf Theorem}\cite{nrt92, k92, k94}. {\it For the generalized
quantum dimensions $Q_{\ell m}$ one finds the following relations}
 \beq \frac{1}{L(1)} \sum_{\ell =1}^k
L\left(\frac{1}{Q_{\ell m}^2}\right) = \frac{3k}{k+2} - 24
\Delta_m^{(k)} +6m. \lleq{nrt}

For $m=0$ this theorem reduces to the central charge result in
terms of the quantum dimensions $Q_{\ell} = S_{\ell 0}/S_{00}$
\begin{equation} \frac{1}{L(1)} \sum_{\ell=1}^k
L\left(\frac{1}{Q_{\ell}^2}\right) = \frac{3k}{k+2}.
\end{equation}
that was obtained earlier in \cite{kr87,l81}.

It follows that the quantum dimensions contain the essential
information about the spectrum of the conformal field theory and
Rogers' dilogarithm provides, via Euler's dilogarithm, the map
from the quantum dimensions to the central charge and the
anomalous dimensions. A review of these results and references to
the original literature can be found in \cite{k94}.

In the case of the elliptic curve $E_4$ the exactly solvable model
is a tensor product two SU(2) theories at conformal level $k=2$,
equipped with the diagonal affine invariant \beq \mathP_{(1,1,2)}
\supset E_4~~~\sim ~~~\left(\rmSU(2)_{k=2,A}^{\otimes
2}\right)_{\rmGSO}. \eeq The quantum dimensions of the SU(2)
theory at level $k=2$ take values in the quadratic extension
$\mathQ(\sqrt{2})$ and therefore we find that the field of quantum
dimensions provides a physical explanation of the emergence of the
Legendre character in the modularity relation
(\ref{quartic-modularity}). Hence the main result of this paper
can be interpreted as a geometric derivation of a conformal field
theoretic object $-$ the conformal field theoretic modular form
$\Theta^2_{1,1}(4\tau)^2$ is the twist of the geometric modular
form defined by the Hasse-Weil L-function via the quadratic
character associated to the number field of the quantum
dimensions.

This physical explanation of the origin of the twist makes it
clear why no character was necessary in the discussion in
\cite{su03} of the plane cubic elliptic curve $C_3$ $-$ the
quantum dimensions of the affine theory $A_1^{(1)}$ at conformal
level $k=1$ take values in the field of rational number $\mathQ$.

\vskip .3truein

\section{Generalizations}

For more general curves of higher genus it is possible to
decompose the Jacobians in a way similar to the analysis in
previous sections. In general the factorization behavior can be
quite involved and in the following we briefly describe what is
known about the splitting behavior. For more detail we refer to
the original literature \cite{f61, r78} (see also \cite{lps03,
lss04}). The reason why this factorization behavior is relevant in
the present context is that the splitting occurs on the level of
isogenies, a map between abelian varieties that is weaker than an
isomorphism. It is however known that isogeneous varieties have
the same L-functions. Also important is that the varieties that
emerge in this factorization admit complex multiplication and
therefore their L-functions are determined by algebraic Hecke
characters (see e.g. \cite{lps03, lss04}). Hecke L-functions are
known to be modular and therefore the factorization of Jacobian
varieties allows to determine modular forms associated to these
higher genus curves.

It was shown by Faddeev  \cite{f61}\fnote{4}{More accessible
references are \cite{w76} \cite{g78}, \cite{r78}.} that the
Jacobian variety $J(C_d)$ of Fermat curves $C_d\subset \mathP_2$
of prime degree splits into a product of abelian factors
$A_{\cO_i}$ \beq J(C_d) \cong \prod_{\cO_i \in
\cI/(\mathZ/d\mathZ)^{\times}} A_{\cO_i}, \eeq where the set $\cI$
provides a parametrization of the cohomology of $C_d$, and the
sets $\cO_i$ are orbits in $\cI$ of the multiplicative subgroup
$(\mathZ/d\mathZ)^{\times}$ of the group $\mathZ/d\mathZ$. More
precisely it was shown that there is an isogeny \beq i: J(C_d)
\lra \prod_{\cO_i \in \cI/(\mathZ/d\mathZ)^{\times}}
A_{\cO_i},\eeq where an isogeny $i: A \ra B$ between abelian
varieties is defined to be a surjective homomorphism with finite
kernel. Explicitly, $\cI$ is the set of triplets $(r,s,t)$
parametrizing a basis of the cohomology \beq \rmH^1(C_d)= \left\{
\om_{r,s,t} = x^{r-1}y^{s-d}dx~|~r,s,t \in \mathN,
0<r,s,t<d,~r+s+t=0~\rmmod~d \right\}, \eeq and the abelian
varieties $A_d^{[(r,s,t)]}$ are associated to orbits $[(r,s,t)]$
of the triplets $(r,s,t)$ with respect to the group
$(\mathZ/d\mathZ)^{\times}$.

The periods of the Fermat curve have been computed by Rohrlich
\cite{r78} to be \beq \int_{\cA^j\cB^k\kappa} \om_{r,s,t} =
\frac{1}{d} B\left(\frac{s}{d},\frac{t}{d}\right)
(1-\xi^s)(1-\xi^t)\xi^{js+kt}, \eeq where $\xi$ is a primitive
$d-$th root of unity, and \beq B(u,v) = \int_0^1
t^{u-1}(1-v)^{v-1}dt \eeq is the classical beta function.
$\cA,\cB$ are the two automorphism generators
\bea \cA(1,y,z) &=& (1, \xi y, z) \nn \\
\cB(1,y,z) &=& (1, y,\xi z)\eea and $\kappa$ is the generator of
$\rmH_1(C_d)$ as a cyclic module over $\mathZ[\cA,\cB]$. The
period lattice of the Fermat curve therefore is the span of \beq
\left(\dots, \xi^{jr+ks}(1-\xi^r)(1-\xi^s) \frac{1}{d}
B\left(\frac{r}{d},\frac{s}{d}\right), \dots
\right)_{\stackrel{1\leq r,s,t \leq d-1}{r+s+t=d}},~~ \forall
0\leq j,k\leq d-1. \eeq

The abelian factor $A_{[(r,s,t)]}$ associated to the orbit
$\cO_{r,s,t}=[(r,s,t)]$ can be obtained as the quotient \beq
A_{[(r,s,t)]} = \mathC^{\vphi(d_0)/2}/\Lambda_{r,s,t}, \eeq where
$d_0 = d/\rmgcd(r,s,t)$ and the lattice $\Lambda_{r,s,t}$ is
generated by elements of the form \beq
\si_a(z)(1-\xi^{as})(1-\xi^{at})
\frac{1}{d}B\left(\frac{<as>}{d},\frac{<at>}{d}\right), \eeq where
$z\in \mathZ[\mu_{d_0}]$, $\si_a \in
\rmGal(\mathQ(\mu_{d_0})/\mathQ)$ runs through subgroups of the
Galois group of the cyclotomic field $\mathQ(\mu_{d_0})$ and $<x>$
is the smallest integer $0\leq x <1$ congruent to $x$ mod $d$.

Alternatively, the abelian variety $A_d^{r,s,t}$ can be
constructed in a more geometric way as follows. Consider the
orbifold of the Fermat curve $C_d$ with respect to the group
defined as \beq G_d^{r,s,t} = \left\{(\xi_1,\xi_2,\xi_3)\in
\mu_d^3~|~ \xi_1^r\xi_2^s\xi_3^t=1 \right\}. \eeq The quotient
$C_d/G_d^{r,s,t}$ can be described algebraically
via projections \bea T_d^{r,s,t}: C_d &\lra &C_d^{r,s,t} \nn \\
(x,y) &\mapsto & (x^d, x^ry^s) =:(u,v), \eea which map $C_d$ into
the curves \beq C_d^{r,s,t} = \left\{v^d = u^r(1-u)^s\right\}.
\eeq

For prime degrees the abelian varieties $A_d^{r,s,t}$ can be
defined simply as the Jacobians $J(C_d^{r,s,t})$ of the
projections $C_d^{r,s,t}$. When $d$ has nontrivial divisors $m|d$,
this definition must be modified as follows. Consider the
projected Fermat curves \bea C_d &\lra & C_m \nn \\ (x,y) &\mapsto
&(\bx, \by):= \left(x^{\frac{d}{m}},y^{\frac{d}{m}}\right), \eea
whose Jacobians can be embedded as $e: J(C_m) \lra J(C_d)$.
Composing the projection $T_d^{r,s,t}$ as \beq J(C_m)
\stackrel{e}{\lra} J(C_d) \stackrel{T_d^{r,s,t}}{\lra}
J(C_d^{r,s,t}) \eeq for all proper divisors $m|d$ leads to a
collection of subvarieties $\cup_{m|d} T_d^{r,s,t}(e(J(C_m)))$.
The abelian variety of interest then is defined as \beq
A_d^{r,s,t} = J(C_d^{r,s,t})/\cup_{m|d}T_d^{r,s,t}(e(J(C_m))).
\eeq

The abelian varieties $A_d^{r,s,t}$ are not necessarily simple but
it can happen that they in turn can be factored. This question can
be analyzed via a criterion of Shimura-Taniyama, described in
\cite{st61}. Applied to the $A_d^{r,s,t}$ discussed here the
Shimura-Taniyama criterion involves computing for each set
$H_d^{r,s,t}$ defined as \beq \rmH_d^{r,s,t} := \left\{a\in
(\mathZ/d\mathZ)^{\times}~|~<ar>+<as>+<at>=d\right\} \eeq another
set $W_d^{r,s,t}$ defined as \beq W_d^{r,s,t} = \left\{a \in
(\mathZ/d\mathZ)^{\times}~|~ aH_d^{r,s,t} = H_d^{r,s,t} \right\}.
\eeq If the order $|W_d^{r,s,t}|$ of $W_d^{r,s,t}$ is unity then
the abelian variety $A_d^{r,s,t}$ is simple, otherwise it splits
into $|W_d^{r,s,t}|$ factors \cite{kr78}.

\vskip .3truein

{\bf ACKNOWLEDGEMENT.}

Part of this work was completed while RS was supported as a
Scholar at the Kavli Institute for Theoretical Physics in Santa
Barbara. It is a pleasure to thank the KITP for hospitality. This
work was supported in part by the National Science Foundation
under Grant No. PHY99-07949, an IUSB Faculty Research Grant (ML),
and a KSU Incentive Grant for Scholarship (RS). Most of the
results of this work have been presented in talks at the Banff
Institute Research Station in December 2003, and the Schr\"odinger
Institute in May 2004. RS is grateful to these institutions for
hospitality.

\end{document}